\documentstyle[12pt]{article}
\newcommand{\beq}{\begin{equation}}
\newcommand{\eeq}{\end{equation}}
\newcommand{\bra}{\begin{array}}
\newcommand{\era}{\end{array}}
\newcommand{\te}{\theta}
\newcommand{\be}{\beta}

\newcommand{\de}{\delta}

\newcommand{\ep}{\epsilon}
\title{Spin-Magnetic Field Interaction and Realization of Fractional Supersymmetry  }  
\author{J.Douari$^{1}$ and E. H EL Kinani$^{2}$}        
\begin{document}        
\maketitle                
D\'epartement de Physique, L.P.T, Facult\'e des Sciences, Universit\'e $M^{ed} V$, Rabat, B.P 1014, Morocco. \\
\vskip 2cm

{\bf {Abstract}}.\\

The fractional supersymmetry in the case of the non-relativistic motion of one anyon with fractional spin is realized. Thus the associated Hamiltonian is discussed.

\vskip 10cm

\hrule 
$^{(1)}$E-mail: douarij@ictp.trieste.it\\
$^{(2)}$Permanent address: D\'epartement de Math\'ematique, Facult\'e des Sciences et Technique, Boutalamine B.P 509,  Errachidia, Morocco.\\


\newpage
\section{Introduction}
\hspace{.3in}Many studies have been carried on the anyon which interpolates between boson and fermion. This kind of particle is characterized by fractional charge, spin and statistics$^{\cite{1,2,3}}$. Thus, owing Wilczek the anyon is defined as charge-flux composite. Where the particle moves in the magnetic field $B$ created by an infinitely long and thin solenoid along the $x_{3}$-axis.\\

Otherwise, the issue of fractional supersymmetric (FSUSY) quantum mechanics has received considerable attention in the literature$^{\cite{4,5,6,7}}$. Which involves the use of the generalized Grassmann variable $\te$ of order $f$ satisfying $\te^{f}=0$.\\

In fact our work here is concerned to investigate the possibility of the existance of the FSUSY at the presence of the fractional spin-mangetic field interaction . Precisely, in this letter, we indicate that the introduction of the potential energy due to the existance of the spin $S = \frac{1}{f}-P$, (where P is a projector) lead to the realization of the FSUSY.\\
\section{Anyon's Hamiltonian}
\hspace{.3in}In two space dimensions there are a specific particles described by a fractional statistics. These particles are called anyons and defined by Wilczek$^{\cite{2}}$ as a coupling of its charge $q$ and one pure chern-Simons gauge field $\bf{A}\rm$. With $\bf{A}\rm$ is considered the vector potential for an infinity long and thin solenoid passing through the origin and directed along the $x_{3}$-axis.\\

A collection of $N$ point particles moving non-relativistically on a plane is described by the following Lagrangian
\begin{equation}
L=L_m +L_{CS}+L_{int},
\end{equation}
where
\begin{equation}
L_m =\sum\limits_{i=1}^{N}\frac{1}{2}m_i \dot{x}^{2}_{i}
\end{equation}
is the matter Lagrangian, while
\begin{equation}
L_{CS}=\frac{\kappa}{2}\int d^2 \bf{x}\rm\epsilon^{\alpha\beta\gamma}\partial_{\alpha}A_{\beta}A_{\gamma}
\end{equation}
is the Chern-Simons Lagrangian, and
\beq
L_{int}=\frac{1}{c}\sum\limits_{i=1}^{N}e_{i}\bf{x}\rm_{i}\cdot\bf{A}\rm (t,\bf{x}\rm _{i})-\sum\limits_{i=1}^{N}e_{i}A^{0}(t,\bf{x}\rm _{i})
\eeq
is the interaction Lagrangian. In these expressions $m_i$ is the particular mass and $\dot{x}$ is the time derivative of the position vector $\bf{x}\rm_i $ (the particle velocity), $e_i$ the particle charge, $c$ is the velocity of light, and $\kappa$ is a measure of the interaction.\\

Let us consider now a non-relativistic particle of mass $m$ and electric charge $q$ moving in the magnetic field $B$ created by the proposed solenoid. In the appropriate chosen system of units $c=m=q=\hbar=1$ and in neglecting the motion along the solenoid, the dynamics will be in the $(x_{1},x_{2})$-plan, the Hamiltonian can be written as follows
\begin{equation}
H=-\frac{1}{2}(\nabla-i\bf{A}\rm)^{2},
\end{equation}
where
\beq
\bra{rl}
A^{i}=\frac{\phi}{2\pi}\ep_{ij}\frac{x^{j}}{|x|^{2}},& i=1,2
\era
\eeq
with $\phi$ is the flux of the solenoid and $(\ep_{ij})=\pmatrix{0&1\cr -1&0\cr}$. In this case the magnetic field $B$ is
\beq
B=\nabla\wedge\bf{A}\rm=\phi\de^{(2)}(x)
\eeq
i.e. the field $B$ is localized at the origin. Yet, the Hamiltonian $(5)$ can be written in the complex notation. Let us now introduce the following complex coordinates
\beq
\bra{llllllllll}
z= x^{1}+ix^{2},& \partial=\frac{1}{2}(\partial_{1}-i\partial_{2})\\
\\
\bar{z} =x^{1}-ix^{2},& \bar{\partial}=\frac{1}{2}(\partial_{1}+i\partial_{2})\\ \\
\partial_{1}\equiv\frac{\partial}{\partial x^{1}},& \partial\equiv\frac{\partial}{\partial z}\\ \\
\partial_{2}\equiv\frac{\partial}{\partial x^{2}},& \bar{\partial}\equiv\frac{\partial}{\partial \bar{z}}\\ \\
A=-\frac{1}{2}(A^{1}-iA^{2})=-i\frac{\phi}{4\pi}\frac{1}{z}\\ \\
\bar{A} =-\frac{1}{2}(A^{1}+iA^{2})=i\frac{\phi}{4\pi}\frac{1}{\bar{z}}.
\era
\eeq
In a straightforward calculation, we can obtain the following commuattion relations those the elements of $(8)$ satisfy
\beq
\bra{lllllll}
\lbrack z, \partial\rbrack =1,& \lbrack \bar{z}, \bar{\partial}\rbrack =1,& \lbrack \bar{z}, \partial\rbrack =0=\lbrack z, \bar{\partial}\rbrack \\
\\
\lbrack \frac{1}{z}, \partial\rbrack =-\frac{1}{z^2}, \lbrack \frac{1}{\bar{z}}, \bar{\partial}\rbrack =-\frac{1}{\bar{z}^2},\\
\lbrack \frac{1}{z}, \bar{\partial}\rbrack =-\pi\de^{(2)}(z)=\lbrack \frac{1}{\bar{z}}, \partial\rbrack .
\era
\eeq
Then, owing to the equations $(8)$ and $(9)$ the Hamiltonian $(5)$ can take a new form in terms of complex coordinates
\beq
H=-2\partial\bar{\partial}-\frac{\phi^{2}}{8\pi^{2}}\frac{1}{z\bar{z}}.
\eeq
Always in the complex notation let us define the operators $\pi$ and $\pi^{+}$ as follows
\beq
\bra{rcl}
\pi=2(i\bar{\partial}+\bar{A})
\pi^{+}=2(i\partial+A).
\era
\eeq
These two operators satisfy
\beq
\lbrack \pi,\pi^{+}\rbrack =2B.
\eeq
\hspace*{.3in}Again once, in using the operators of $(11)$ it is very easy to verify that the Hamiltonian $(5)$ can be rewritten as
\beq
H=\frac{1}{4}(\pi\pi^{+}+\pi^{+}\pi).
\eeq
\section{Fractional Supersymmetry}
\hspace{.3in}Let us mention here that the Hamiltonian (13) corresponds to that discussed in the reference \cite{9}, where the authors presented an exactly solvable model of anyons which have an attractive hard-core interaction, and they considered their Hamiltonian into an f=2 supersymmetric form.\\  

In this work, an important fact we would like to shed light on is that the fractional spin maghnetic field interaction lead to the realization of FSUSY quantum mechanics. To begin, let us assume that there are one interaction between the spin $S$ of our particle and the associated magnetic field $B$. In this case, an additional potential energy written as
\beq
H_{I}=S.B
\eeq
can be existed. Consequently, the Hamiltonian will be
\beq
H=\frac{1}{4}(\pi\pi^{+}+\pi^{+}\pi)+S.B.
\eeq
\hspace*{.3in}In order to construct the FSUSY in this case, let us recall the basic notations of the generalized Grassmanian variables $\te$. Such variables of order $f$, $f=2,3,...$ and their derivatives $\partial_{\te}$ satisfy the generalized commutation relation
\beq
\lbrack \partial_{\te},\te\rbrack_{q}\equiv\partial_{\te}\te-q\te\partial_{\te}=\be(1-q),
\eeq
where $\be$ is an arbitrary parameter and $q\in\bf{C}\rm$ a primitive $f^{th}$ root of unity $q^{f}=1$, and $\te^{f}=0=\partial_{\te}^{f}$ for a given $f$. Thus a matrix realization of $\te$ and $\partial_{\te}$ is given by
\beq
\bra{cc}
\theta= \pmatrix{0&a_{1}&0&0&0\cr
0&0&a_{2}&0&0\cr
0&0&0&\ddots&0\cr
0&0&0&0&a_{F-1}\cr 
0&0&0&0&0\cr}\\ \\
\partial_{\theta}= \pmatrix{0&0&0&0&0\cr 
b_{1}&0&0&0&0\cr 
0&b_{2}&0&0&0\cr 
0&0&\ddots&0&0\cr 
0&0&0&b_{F-1}&0\cr}
\era
\eeq
with the constraint $a_{i}b_{i}=\be(1-q^{-i})$.\\

Yet, we suppose that the anyon is characterized by the fractional spin given by
\beq
\bra{rl}
S=\frac{1}{f}-P,& f=2,3,...
\era
\eeq
where $P$ is projector operator expressed in terms of $\te$ and $\partial_{\te}$ as follows
\beq
\bra{rcl}
P=a \te^{f-1} \partial_{\te}^{f-1},& P^{2}=P,& a=\frac{1}{f \beta^{f-1}}.
\era
\eeq
\hspace*{.3in}Furthermore, to construct the $N=1$ FSUSY we can introduce the folowing fractional superchrage
\begin{equation}
Q= \partial_{\theta}^{f-1}{\pi}+a \theta {\pi^{+}}+(1-P)\theta
\end{equation}

One can easily check that the above supercharge satisfy the following relations

\begin{equation}
Q^{f}=H  \:\:\:\:\:\:\,\:\:\:\:\:\:\ [\ H, Q]=0, \:\:\:\ f=2, 3,...
\end{equation}
\\
which define the FSUSY quantum mechanics of order $f$.\\

An interesting case is obtained when $f=2$, in this situation we reproduce the Pauli Hamiltonian \cite{8}
\begin{equation}
H= {\frac{1}{2}}((\partial_{1}+ A_{1})^{2}+(\partial_{2}+ A_{2})^{2}) -{\frac{1}{2}}{\sigma}.B.
\end{equation}
\\
Then the supercharge and spin are given by the following formulas 
\begin{equation}
Q=2(\sigma_{1}(P_{1}-A_{1})+\sigma_{2}(P_{2}-A_{2}))\:\:\:\:\, \:\:\:\:\ S= {\frac{-1}{2}}{\sigma}_{3},
\end{equation}
\\
where we have used the choise $\beta =\frac{1}{2}$ $(a=1)$, $\theta =  \sigma_{+}$, $\partial_{\theta}= \sigma_{-}$, with  $ \sigma_{\pm}=\sigma_{1} \pm i \sigma_{2}$ and finally $\sigma =(\sigma_{1}, \sigma_{2}, \sigma_{3})$ are the Pauli matrices.\\

In the generalized case, using the realization Eq(17) in terms of $f \times f$ matrices, the projector P is found$^{\cite{10}}$ to be a matrix whose upper left entry is one and all others zero. Therefore, the matrix S is $S \equiv diag(\frac{f-1}{f}, \frac{-1}{f}, \frac{-1}{f},...)$, then $s=\frac{f-1}{f}$ is a (non-degenerate) eigenvalues of S and $s=\frac{-1}{f}$ is $f-1$ (degenerate). \\

In conclusion, we have shown how some fractional spin-magnetic field interaction lead to the realization of fractional supersymmetry. It is interesting to investigate the consequence of this new symmetry.\\


\begin{thebibliography}{21}
\bibitem{1}J. M. Leinaas and J. Myrheim, Nuovo Cimento B\bf{37}, \rm{1} (1977).
\bibitem{2}F. Wilczek. Phys. Rev. Lett. \bf{48}, \rm{1144} (1982).
\bibitem{3}R. Jackiw, Phys. Rev. D\bf{42}, \rm{3500} (1990).
\bibitem{4}S. Durand, M. Mayrand, V. P. Spiridonov and L. Vinet, Mod. Phys. Lett. {\bf{A6}}, 3163 (1991).
\bibitem{5}M. Rausch de Traubenberg and P. Simon, hep-th/9606188; Nucl. Phys. {\bf{B482}}, 325(1996); and refs therein.
\bibitem{6}J. A. de Azc\'arraga and A. J. Macfarlane, J. Math. Phys. \bf{37}\rm, 1115 (1996).
\bibitem{7}R. S. Dunne, preprint hep-th.9703111.
\bibitem{8}V. M. Tkachul and S. I. Vakarchul, Phys. Lett. {\bf{A228}}, 141 (1997).
\bibitem{9}S. M. Girvin, A. H. Mac Donald, M. P. A. Fisher, S. 6J. Rey and J. P. Sethna, Phys. Rev. Lett. \bf{65}, \rm{1671} (1990).
\bibitem{10}S. Durand, Phys. Lett. \bf{A7}, \rm{2905} (1992); \bf{A8}, \rm{2323} (1993); \bf{A8}, \rm{1195} (1993).
\end{thebibliography}
\end{document}